%Paper: hep-ph/9305280
%From: RINGWALD@crnvma.cern.ch
%Date: Tue, 18 May 93 10:14:09 SET

%%%%%%%%%%%%%%%%%%%%%%%%%%%%%%%%%%%%%%%%%%%%%%%%%%%%%%%%%%%%%%%%%%%%%%%%%
\magnification=1200
\font\titlefnt=cmbx10 scaled \magstep2
\baselineskip=6mm

$$\eqno{\rm
CERN-TH.6862/93}$$
\vskip 1truecm
\centerline{\titlefnt Multi-W(Z) Production in}
\centerline{\titlefnt High-Energy Collisions?}
\vskip 1.5truecm

\centerline{A. Ringwald}
\centerline{CERN, Geneva}
\centerline{Switzerland}

\vskip 3truecm
\centerline{ABSTRACT}
\noindent
There exists the possibility that
the cross-section for the
nonperturbative
production of many, ${\cal O}(\alpha_W^{-1} \simeq 30
)$,
weak gauge bosons may be as large as ${\cal O}(100\ {\rm pb}\
-\  10\ \mu$b$)$ above a parton-parton center-of-mass threshold in
the range $2.4\ -\ 30$ TeV.
We review the theoretical considerations
which lead to this suggestion and outline its phenomenological
implications, both for present
cosmic ray as well as for future
collider experiments.

\vskip 2truecm
\centerline{\it Lecture presented
at the
4th Hellenic School on Elementary Particle Physics}
\centerline{\it September 2-20, 1992, Corfu, Greece}
\vskip 1.5truecm

$$\leqno{\rm
CERN-TH.6862/93}$$
$$\leqno{\rm April\ 1993}$$
\nopagenumbers
\vfil\eject

\nopagenumbers
\null\vfill\eject
\advance\pageno -2

\leftline{\titlefnt 1. Introduction}
\bigskip\noindent
Experiments at LEP are vindicating triumphantly the Standard Model,
testing it with a precision approaching one part in a thousand.
The only missing links are the top quark and the Higgs boson.
So far LEP has given us no direct evidence for physics beyond
the Standard Model.
Indeed,
the Standard Model may be
valid, as an effective theory, up
to very high energies, say the Planck mass, 10$^{19}$ GeV, if
the Higgs mass is below  several hundreds of GeV.
\par\noindent
This, however, does not necessarily
imply that no new phenomena will be seen
in the multi-TeV range, i.e. in
the energy range which will be explored by
LHC
and SSC.
There is the intriguing
possibility [1-6]
that,
above a parton-parton center-of-mass (CM)
threshold in
the range $2.4\ -\ 30$ TeV,
the cross-section for the
nonperturbative
production of many, ${\cal O}(\alpha_W^{-1} \simeq 30
)$,
weak gauge bosons may be as large as ${\cal O}(100\ {\rm pb}\
-\  10\ \mu$b$)$. Unfortunately, there is only circumstantial
evidence for this to happen which is, to a large extent, just
based on the observation that lowest-order
calculations for the production of ${\cal O}(\alpha_W^{-1})$
weak gauge bosons, both with and without association of baryon and
lepton number violation,
violate unitarity near the threshold
${\cal O}(m_W/\alpha_W)$.
\par\noindent
It is the purpose of this lecture
to review the theoretical status of the subject
and to discuss prospects
of cosmic ray and collider experiments
 to constrain
or even observe multi-W(Z) production in high-energy collisions.

\bigskip
\leftline{\titlefnt 2. `Theory' of Multi-W(Z) Production}
\bigskip
\leftline{\bf 2.a. Multi-W's(Z's) with  B\ \& L Violation}
\par\noindent
That there may be a large cross section for the production
of ${\cal O}(\alpha_W^{-1})$ weak gauge bosons was
suggested first
[1-3]
by investigations of electroweak baryon (B) and
lepton (L) number violation.
For definiteness,
we will consider
the prototype model of weak interactions,
the
fundamental SU(2) Higgs model with chiral fermions,
defined by the following action:
$$S\lbrack W,\phi ,\overline{\Psi}_L, \Psi_L\rbrack
\equiv$$
$$
\int d^4 x\
\biggl\{ -{1\over 2}{\rm tr}\bigl( F_{\mu\nu}F^{\mu\nu} \bigr) +
 \mid D_\mu\phi\mid^2 -\lambda \Bigl(
\mid\phi\mid^2-{v^2\over 2}\Bigr)^2 +
\sum_{j=1}^{12} \overline{\Psi}_L^{(j)}i\gamma_\mu D^\mu
\Psi_L^{(j)}
\biggr\}.
\eqno(1)$$
This model corresponds to the
electroweak theory
in the limit of
$\sin^2\theta_W\to 0$ and of vanishing fermion masses.
The superscripts at the fermion   fields $\Psi^{(j)}$,
$j=1,...,12$, label the different fermionic flavours
in the Standard Model with three generations.
\par\noindent
Due to the chiral anomaly [7],
B  and L are not strictly conserved
in the Standard Model [8].
In the presence of non-trivial SU(2) gauge fields $W_i$,
the fermionic quantum
numbers change according to  [8]
$$
\triangle {\rm L}_e
=
\triangle {\rm L}_\mu
=
\triangle {\rm L}_\tau
=                     {1\over 3}
\triangle {\rm B}
=  -\triangle N_{\rm CS}
,\eqno(2)$$
where
$$N_{\rm CS} \equiv {\alpha_W\over 4\pi}
\int d^3x\ \epsilon^{ijk} {\rm tr} \biggl(
F_{ij} W_k - {2ig\over 3} W_iW_jW_k\biggr) \eqno(3)$$
denotes the Chern-Simons number of the gauge field.
As is suggested by eqs. (2) and (3), one needs strong, nonperturbative
 gauge
fields, of order $g^{-1}$, in the intermediate state in order
to change the Chern-Simons number, or, equivalently, the
fermion numbers by an integer amount. This is reflected by the
fact that there exists an energy barrier [9]
between gauge fields
whose Chern-Simons numbers differ by an integer (Fig. 1).
The minimum barrier height is given by the energy of
a static
saddle-point solution, the so called ``sphaleron'' [10], which
slightly depends on the Higgs mass and is of order
$$M_{\rm sp} = B(m_H/m_W)\ \pi {m_W\over \alpha_W}
\simeq ( 7-14 )\ {\rm TeV} .\eqno(4)$$
At low energies ($\ll M_{\rm sp}$)
anomalous B\ \& L violating processes are
only possible by quantum tunneling, i.e. the corresponding
amplitudes
are exponentially suppressed by a Gamow factor,
$$
{\cal A}^{\triangle N_{\rm CS} =1}_{\mid
E\ll M_{\rm sp}}
\propto {\rm e}^{-{2\pi\over \alpha_W}}
\sim 10^{-78} ,\eqno(5)$$
which leads to unobservably small cross sections or decay
rates [8].
\par\noindent
Let us consider now
high-energy parton-parton (e.g.
quark-quark or neutrino-quark)
collisions.
As has been suggested in refs. [11,12], one expects that the dominant
B\ \& L violating processes will involve ${\cal O}(\alpha_W^{-1})$
W's (Z's), simply because sphaleron-like intermediate states will
typically decay into many W's and Z's [13].
Therefore one should
try to calculate anomalous amplitudes involving an arbitrary number
of weak gauge bosons, which are given, up to analytic continuation
and LSZ reduction, by the following euclidean
path integral:
$$
{\cal A}^{\triangle N_{\rm CS} =1}_{n_W}
\sim
\int {\cal D}W{\cal D}\phi{\cal D}\Psi_L{\cal D}\overline{\Psi}_L\
{\rm e}^{ -S_E [
W,\phi ,\overline{\Psi }_L ,\Psi_L ]}
\prod_{i=1}^{12}
\overline{\Psi}^{(i)}_L (x_i) \prod_{j=1}^{n_W}
W_{\mu_j}(y_j)
 .\eqno(6)$$
The leading
contribution
in $\alpha_W$
to the path integral appearing in
eq. (6)
may be found by semiclassical methods:
The integral receives its dominant contribution
from that region in field space
in which
the euclidean action, $S_E$,
attains its minimum; i.e. one has
to find a
classical solution,
with $\triangle N_{\rm CS}=1$, and
expand the integrand about it. This classical solution\footnote*{
Actually, some complication arises in the electroweak theory
due to its mass scale introduced by symmetry breaking:
by scaling arguments one may show that strictly speaking
no exact solution exists [15].
A systematic
semiclassical expansion can be done nevertheless if
the field trajectory about which one expands is
a `constrained instanton' [15] or, more generally, a `valley trajectory'
[16].}
is called `instanton' [14].

\bigskip\vskip 6truecm
\noindent
{\bf Fig. 1}:
{\it The static bosonic energy $E$
versus the Chern-Simons number and
the effective radius $R$ of the energy density distribution.
Gauge fields with integer difference in Chern-Simons number are
separated by an energy barrier. The minimum barrier height is taken
at the sphaleron, a saddle-point solution of the static field equation,
with an energy $M_{\rm sp}\sim m_W/\alpha_W$ and a radius
$\sim m_W^{-1}$.}
\bigskip\noindent
Schematically,
one obtains [1,2], in the leading-order expansion about
the instanton,
$${\cal A}^{\triangle N_{\rm CS} = 1}_{n_W\ {\rm lead.-ord.}}
\sim
n_W!\  \alpha_W^{n_W/2}\
{\rm e}^{-2\pi /\alpha_W }
\ m_W^{-n_W}
 .\eqno(7)$$
At high energies,
 these point-like amplitudes violate
unitarity, since according to eq. (7)
the corresponding cross-sections grow like
phase space.
Due to the factorial growth of the amplitudes (7) with the number
of produced gauge bosons, this
violation of unitarity sets in already at
multiplicities of order $n_W^0\sim 1/\alpha_W$ and at
parton-parton CM
energies of
order $\sqrt{\hat s_0}\sim m_W/\alpha_W$.
\par\noindent
A violation of unitarity is, of course, unacceptable
and indicates the importance of higher-order corrections.
There are strong arguments that the corrections to the
fixed-multiplicity amplitudes exponentiate in the total cross-section
of B\ \& L violation, such that, to exponential accuracy, the
latter can be written as [17-19]:
$${\hat \sigma}_{\rm tot}^{\triangle N_{\rm CS}=1}\propto
 \exp\Biggl\lbrack
{4\pi  \over \alpha_W} F
\biggl( {\sqrt{\hat s}\over M_0 } \biggr)
\Biggr\rbrack
 ,
\eqno(8)$$
where
$M_0\equiv \pi m_W/\alpha_W$ is of the order of the sphaleron scale (4).
{}From eqs. (5) or (7) we know that $F(0)=-1$.
The crucial question is, whether the `holy grail function'
$F$ approaches zero at high energies: Only in this case one might
hope that the total cross-section of B\ \& L violation becomes
of observable size.
\bigskip\vskip 6truecm
\noindent
{\bf Fig. 2}:
{\it Guesses for the high-energy behaviour of the holy grail function
F. Solid line, ref. [28]. Dashed line, ref. [29].}
\bigskip
\vfil\eject
\par\noindent
The perturbative expansion
about the instanton yields a low-energy expansion of
$F$, whose first few terms are given by
[18,20-26]
$$F
( \epsilon )
=
-1 +
0.34\cdot
\epsilon^{4/3} -0.09\cdot
\epsilon^2 +
0.01\cdot \biggl( 4- 3{m_H^2\over m_W^2} \biggr)
\cdot
\epsilon^{8/3}
\cdot
\ln \biggl( {1\over \epsilon} \biggr)
+                  $$
$$+
{\cal O}\biggl(
\epsilon^{8/3}
\cdot {\rm const.}
\biggr)
,\eqno(9)$$
where $\epsilon \equiv \sqrt{\hat s}/M_0$.
{}From this result the following conclusions can be drawn:
\item{--} The total cross-section of  B\ \& L violation is
exponentially growing, albeit exponentially small,
at $(m_W\ll ) \sqrt{\hat s}\ll M_0$.
\item{--} The different
terms in the perturbative expansion of the
holy grail function become of comparable size, i.e. the perturbative
expansion breaks down,
at $\sqrt{\hat s}
\sim M_0$.
\par\noindent
As a side-remark we note, that,
at low energies, the total
cross-section (8) is dominated by multi-W(Z) production
($n_W\sim \alpha_W^{-1}$)
rather
than by multi-Higgs production.
\par\noindent
Nothing is known about the behaviour of the holy grail function
at energies of the
order of
the sphaleron scale and above.
There are various guesses
(see Fig. 2) but a systematic calculation of it is
not available at present.
Since instanton-based perturbation theory
breaks down at the sphaleron scale,
new methods have to be
developped in order to attack this problem (for reviews see
ref. [27]) and to decide ultimately if electroweak
B\ \& L violation with the associated production of many W's and
Z's will be observable at LHC or SSC.
\bigskip
\leftline{\bf 2.b. Multi-W's(Z's) without  B\ \& L Violation}
\par\noindent
Soon after the discovery of the high-energy and high-multiplicity
breakdown of perturbation theory about the instanton
it was suggested that also conventional
perturbation theory based on Feynman graphs,
as relevant to processes
without B\ \& L violation, breaks down at
high energies and multiplicities [4,5]. It was argued that
perturbative tree-graph amplitudes for multi-Higgs [4,5]
and multi-W(Z) [4,30]
production have a similar factorial growth with the number of
external legs as the leading-order instanton-induced amplitudes
(7),
$${\cal A}^{\triangle N_{\rm CS} = 0}_{n_W\ {\rm tree-graph}}
\sim
n_W!\ \alpha_W^{n_W/2}\ m_W^{-n_W}
 ,\eqno(10)$$
leading again to a violation of unitarity at large multiplicities
of order $\alpha_W^{-1}$ and energies of order $m_W/\alpha_W$.
Due to the absence of the Gamow factor (5) in B\ \& L conserving
processes, the tree-level
onset of the violation of unitarity
for these processes
may be at a somewhat smaller energy
and multiplicity
than for B\ \& L violating processes.
\par\noindent
There is a lot of ongoing work in this direction, which deals,
for technical reasons, mainly with multi-Higgs production
[31,32] (for a review see [33]).
There are general arguments [34], based on dispersion relations
for forward elastic scattering amplitudes
and the assumption that perturbation theory is asymptotic,
that multi-particle production in
weakly-coupled theories is always
exponentially suppressed,
$${\hat \sigma}_{n_W\geq {\cal O}(\alpha_W^{-1})} \preceq
{\rm e}^{-c/\alpha_W }  .\eqno(11)$$
This,
however, does not exclude the possibility of a small
coefficient $c\ll 1$, such that numerically an
observable cross-section results [35].

\bigskip
\leftline{\titlefnt 3. `Phenomenology'
of Multi-W(Z) Production}
\bigskip\noindent
In this section we want to discuss the prospects to observe
or constrain possible multi-W(Z) phenomena in experiments dedicated
to investigate ultrahigh-energy cosmic rays or in experiments
at future hadron colliders such as LHC or SSC.
\bigskip\noindent
\leftline{\bf 3.a. Working Picture}
\par\noindent
In order to confront the idea of possible multi-W(Z) production
with experiments some
working hypothesis is needed.
We will assume [36],
as suggested by the
results reported above,
a sudden onset
of multi-W(Z)
 phenomena on the parton (quark, lepton) level
above a certain
threshold energy $\sqrt{\hat s_0}$:
$$\hat \sigma_{n_W^0} = \hat \sigma_0 \cdot
\theta
\bigl( \sqrt{\hat s} -\sqrt{\hat s_0 } \bigr) ,
\eqno(12)$$
where the parton cross-section
 $\hat \sigma_0$
and the threshold
$\sqrt{\hat s_0}$ lie in between
$$0.1\ {\rm nb}\ \sim {\alpha_W^2\over m_W^2}\leq
\hat \sigma_0\leq \sigma_{\rm pp}^{\rm inelastic}\cdot
\biggl( {1\ {\rm GeV}\over m_W}\biggr)^2\sim
10\ \mu{\rm b}\ ,
\eqno(13)$$
$$2.4\ {\rm TeV}\ \sim {m_W\over \alpha_W}\leq
\sqrt{\hat s_0}\leq 30\ {\rm TeV} .\eqno(14)$$
For definiteness, we will take
$$n_W^0= 30   \eqno(15)$$
for the number of produced weak
vector bosons\footnote*{Results for other
values of $n_W^0$ may be obtained essentially by scaling [36].}.
\par\noindent
The
cross-section  for
multi-W(Z) production in proton-proton and neutrino-nucleon
collisions (Fig. 3)
is obtained by folding eq. (12) with the corresponding
valence- and sea-quark distributions inside the nucleons:
$$\sigma^{\rm pp}_{n_W^0} (\sqrt{s} )=
\sum_{ij({\rm no\  gluons})}
{1\over 1+\delta_{ij}} \int dx_1\ dx_2\ f_i (x_1 ) f_j (x_2 )\
{\hat \sigma}_{n_W^0} (\sqrt{x_1x_2s})  ,\eqno(16{\rm a})$$
$$\sigma^{\nu {\rm N}}_{n_W^0} (\sqrt{s} )=
\sum_{i({\rm no\ gluons})}
 \int dx\ f_i (x   ) \
{\hat \sigma}_{n_W^0} (\sqrt{xs})  .\eqno(16{\rm b})$$
\bigskip\vskip 6truecm
\noindent
{\bf Fig. 3}:
{\it Universal curves parametrizing the production cross-sections
of $n_W^0$
weak vector bosons in proton-proton (pp) and neutrino-nucleon
($\nu$N) collisions, where $\sqrt{s}$ is the total CM energy (from
ref. [38]).}
\bigskip\noindent
With the help of (16a) it is possible to calculate the expected
event rate of multi-W(Z) processes at LHC ($\sqrt{s}=16$ TeV;
{\cal L}=10$^{34}$ cm$^{-2}$s$^{-1}$)
and
SSC ($\sqrt{s}=40$ TeV;
{\cal L}=10$^{33}$ cm$^{-2}$s$^{-1}$),
as a function of
the parton cross-section $\hat \sigma_0$ and
the threshold energy
$\sqrt{\hat s_0}$.
{}From Fig. 4 it is apparent that LHC can cover only a part of
the parameter space, eqs. (13) and (14),
 which we are contemplating, whereas SSC will be much better in this
respect.
\bigskip\noindent
Let us have a closer look on the hard parton-parton subprocess
in which, say, 30 W's are produced. These W's decay immediately
into about
 400 charged hadrons (mainly $\pi^\pm$'s), with
 average transverse momenta  of order
$$p_T^\pi \sim {\cal O}
(m_W/30 )\sim (2-3) \ {\rm GeV}  ,\eqno(17)$$
 and
400 photons (mainly from $\pi^0$'s).
In addition one has about 5
prompt muons
(3 from W decay and 2 from c, b, or $\tau$ decay),
with transverse momenta of order
$$p_T^\mu \sim {\cal O}( m_W/2 )\sim 40\ {\rm GeV} .\eqno(18)$$
Similar numbers of other prompt leptons, like
electrons, positrons
and neutrinos are expected.
It is hard to imagine that any other process in the Standard
Model can mimick such a final state [36].
However, it is not clear if it will be possible to see any
signal of B\ \& L violation in this multi-particle environment
[36,37].
\bigskip\vskip 8truecm
\noindent
{\bf Fig. 4}:
{\it
`Discovery limits' for multi-W(Z) phenomena at pp colliders and
in underground cosmic ray experiments
(taken from refs.
[39,41]).
\par\noindent
To the left of the solid lines labeled `LHC' and `SSC': more than
100 multi-W(Z)
events per year at the corresponding collider.
\par\noindent
To the left of the dashed line labeled `MACRO': more than one
neutrino-initiated
multi-W(Z)
muon bundle passing through MACRO during 10 years of running,
assuming the (revised)
neutrino flux from Stecker et al.
[43].
\par\noindent
Shaded region: Excluded region from Fly's Eye limits [45], assuming
the (revised)
neutrino flux from Stecker et al.
[43].
}
\vfil\eject
\bigskip\noindent
\leftline{\bf 3.b. Multi-W(Z) Phenomena in Cosmic Rays}
\par\noindent
Since LHC and SSC will be operational only in about a decade,
it is worthwhile [38]
to consider also the prospects of observing
or constraining multi-W(Z) phenomena initiated by the interaction of
ultrahigh-energy
cosmic ray particles
with nucleons in the atmosphere or inside the Earth.
The energies
(in the Earth's rest frame)
of the primaries
should exceed
$$E\geq {\hat s_0\over 2m_{\rm p}} \geq 3.3\cdot
10^6\ {\rm GeV} .\eqno(19)$$
One may consider multi-W(Z) processes
initiated by cosmic ray protons and neutrinos:
The corresponding (model-dependent)
fluxes of protons and neutrinos
are shown in Fig. 5.
\bigskip\vskip 6truecm
\noindent
{\bf Fig. 5}:
{\it
Differential fluxes of diffuse
cosmic ray protons and neutrinos.
\par\noindent
Full line: differential flux of neutrinos
from active galactic nuclei as predicted by Stecker et al. (revised)
[43].
\par\noindent
Dashed line: differential flux of protons according to the
constant-mass-composition model of ref. [44].
}
\bigskip\noindent
Multi-W phenomena initiated by cosmic ray
{\it protons}
are plagued [39]
by small rates and poor signatures due to competing
purely
hadronic processes with
${\cal O}(40)$ mb cross-sections.
The primary multi-W(Z) process
takes place inevitably at the top of the
atmosphere and generates an extensive air shower, whose observable
characteristics
at a ground-level air-shower array, like the
MeV-electron
component and the GeV-muon component, ressembles closely
the characteristics of a generic extensive air shower, initiated
by a cosmic ray hadron via strong   interactions.
Moreover, the overall rate of air showers initiated by protons via
multi-W(Z) production
is
a factor of at least (10 $\mu$b / 40 mb)$\sim 10^{-4}$ smaller
than the rate of generic air showers.
A characteristic difference between
proton-initiated multi-W(Z) air showers
and
generic air showers
is seen in the
lateral distribution of the muons with energies above 1
 TeV,
which may be
observed in underground or underwater detectors: The
most energetic muons, which originate from W-decay and have a
high transverse momentum (see eq. (18)), lead to an excess of
muons at distances larger than about 20 m from the shower core.
However, underground (underwater)
detectors of a size
of about
10$^5$ m$^2$ are needed
in order to expect an appreciable rate of proton-initiated
multi-W(Z) muon bundles [38,39].
That means that only the biggest underwater detectors such as
DUMAND or NESTOR
can expect
sizeable rates.
\par \noindent
By contrast, multi-W(Z) processes initiated by ultrahigh-energy
{\it neutrinos}
compete only with relatively small ${\cal O}({\rm nb})$
charged-current reactions.
Moreover, even for the largest, ${\cal O}(10\ \mu{\rm b})$,
cross-section
we contemplate, over 99\% of the primary-neutrino interactions
(for vertically incident neutrinos) take place within the Earth
rather than in the atmosphere.
Due to the large density of rock (or water),
all the hadrons and photons from W-(Z-)decay are
quickly absorbed.
Only the few   (three,
for $n_W^0=30$ W-bosons)
prompt muons
from W-(Z-)decay,
 with energies of order 100 TeV and transverse momenta
of order 40 GeV,
penetrate further,
giving rise eventually
to a multiple muon event
in an underground (underwater)
detector
[38,40].
The most promising
characteristic features
[38,39,41]
of neutrino-initiated multi-W(Z)
muon bundles, distinguishing them from generic atmospheric
muon bundles, are
the large muon
energies, ${\cal O}(100\ {\rm TeV})$,
leading to visible
non-ionization energy
losses within underground
or underwater detectors [42],
and the small
pairwise muon separation, ${\cal O}(20\ {\rm cm})$.
\par\noindent
In order to calculate the expected rate of neutrino-initiated
multi-W(Z) muon bundles in an underground detector one needs
to know the incident flux of neutrinos at ultrahigh energies, eq. (19).
Recent models have predicted a sizeable ultrahigh-energy
neutrino flux
from active galactic nuclei [43], see Fig. 5.
As can be inferred from Fig. 4, MACRO, a big
($72\ {\rm m}\ \times\ 12\ {\rm m}\ \times\ 4.8\ {\rm m}$)
underground detector
situated in Italy
at a depth of 1.4 km
under the Gran Sasso,
can cover already a (small) region of
multi-W(Z) parameter space [39,41]
until LHC or SSC become operational, under
the assumption
that the predictions of the ultrahigh-energy neutrino
flux
in ref. [43]
are correct.
\par\noindent
It is interesting to note that, under the same assumption, already
a part of multi-W(Z) parameter space is excluded [38-41],
see Fig. 4.
The point is that some of the ultrahigh-energy neutrinos
may initiate, via multi-W(Z) production, extensive
air showers,
starting deep in the atmosphere.
The Fly's Eye air-shower
array searched exactly for such a signature [45],
namely
for air showers initiated by weakly interacting
particles (which we will assume are neutrinos)
which
had penetrated more than 3000 g/cm$^2$ of atmosphere before
interacting, thus excluding photons or hadrons as primaries.
{}From the nonobservation of such showers they
deduce upper limits on the flux times cross-section
for primary energies
in the range
$10^8\ {\rm GeV}\leq E_\nu\leq 10^{11}\ {\rm GeV}$;
explicit
parametrizations of the Fly's Eye limits may be found in refs.
[38,46].
The limit
applies only for $\sigma_{\rm tot}^{\nu N}\leq 10\ \mu{\rm b}$, since
the possibility of flux attenuation is neglected.
Using a particular flux model, like the (revised)
Stecker et al. [43] flux of ultrahigh-energy neutrinos from active
galactic nuclei, the Fly's Eye limit translates into an
upper bound on the neutrino-nucleon cross-section and, finally
 [39,41],
into an exluded region in multi-W(Z) parameter space,
see Fig. 4.
\bigskip
\leftline{\titlefnt 4. Summary}
\bigskip
\noindent
Lowest-order perturbative
calculations of the cross-section for the production of
$n_W$ weak gauge bosons
in
parton-parton (e.g.
quark-quark or neutrino-quark) scattering violate unitarity
at parametrically large multiplicities, $n_W^0\sim\alpha_W^{-1}$,
and energies, $\sqrt{\hat s_0}\sim m_W/\alpha_W$.
This happens
both for processes
with and without associated
B\ \& L violation.
At present, it is an open question
 whether
the {\it actual} (--   beyond perturbation theory --)
multi-W(Z) cross-sections become observably large at
such  multiplicities and energies.
New theoretical methods are needed to answer this important question.
\par\noindent
Multi-W(Z) processes at LHC or SSC would be
clearly distinguishible from any other Standard Model process,
due to the   hadronic and leptonic decays of the W's and Z's,
which
 lead to hundreds of charged
hadrons and photons with transverse momenta in the GeV range, and to
tens of
prompt
leptons with transverse momenta of order 40 GeV.
The question whether B\ \& L violation can directly be seen
in such a multi-particle environment requires further investigations.
Due to its higher proton-proton
CM energy, SSC can cover a much larger
range of parton-parton multi-W(Z) threshold
energies and cross-sections than LHC, see Fig. 4.
\par\noindent
Ultrahigh-energy cosmic ray physics is another area where
one could look for
possible multi-W(Z) production.
Current underground detectors such as MACRO
are already sensitive to
neutrino-initiated multi-W(Z) muon bundles, which are
clearly distinguishible from atmospheric muon bundles,
for sufficiently low
parton-parton threshold energies and large cross-sections,
see Fig. 4.
Thus, we have not to wait until LHC or SSC are operating in order
to constrain or even observe multi-W(Z) production in high-energy
collisions.

\bigskip
\leftline{\titlefnt Acknowledgements}
\bigskip
\noindent
I am indebted to V.V. Khoze, D. Morris, F. Schrempp and
C. Wetterich for their collaboration on different parts of the
subject.

\vfil\eject
\leftline{\titlefnt References}
\bigskip
\item{[1]}
 A. Ringwald, {\it Nucl. Phys.} {\bf B330} (1990) 1.
\item{[2]}
 O. Espinosa, {\it Nucl. Phys.} {\bf B343} (1990) 310.
\item{[3]}
 L. McLerran, A. Vainshtein and M. Voloshin,
{\it Phys. Rev.} {\bf D42} (1990) 171.
\item{[4]} J. Cornwall, {\it Phys. Lett.} {\bf B243} (1990) 271.
\item{[5]} H. Goldberg, {\it Phys. Lett.} {\bf B246} (1990) 445.
\item{[6]} A. Ringwald and C. Wetterich, {\it Nucl. Phys.}
{\bf B353} (1991) 303.
\item{[7]}
 S. Adler, {\it Phys. Rev.} {\bf 177} (1969) 2426;
 J. Bell and R. Jackiw, {\it Nuovo Cimento} {\bf 51} (1969) 47;
W.A. Bardeen, {\it Phys. Rev.} {\bf 184} (1969) 1848.
\item{[8]}
 G. 't Hooft, {\it Phys. Rev. Lett.} {\bf 37} (1976) 8;
{\it Phys. Rev.} {\bf D14} (1976) 3432.
\item{[9]} R. Jackiw and C. Rebbi, {\it Phys. Rev. Lett.}
{\bf 37} (1976) 172; C. Callan, R. Dashen and D. Gross,
{\it Phys. Lett.} {\bf B63} (1976) 334.
\item{[10]} N. Manton, {\it Phys. Rev.} {\bf D28} (1983) 2019;
 F. Klinkhamer and  N. Manton,
{\it Phys. Rev.} {\bf D30} (1984) 2212.
\item{[11]} H. Aoyama and H. Goldberg, {\it Phys. Lett.}
{\bf B188} (1987) 506.
\item{[12]} P. Arnold and L. McLerran, {\it Phys. Rev.}
{\bf D37} (1988) 1020.
\item{[13]} M. Hellmund and J. Kripfganz, {\it Nucl. Phys.}
{\bf B373} (1992) 749.
\item{[14]}
 A. Belavin, A. Polyakov, A. Schwarz and  Yu. Tyupkin,
{\it Phys. Lett.} {\bf B59} (1975) 85.
\item{[15]}
 I. Affleck, {\it Nucl. Phys.} {\bf B191} (1981) 445.
\item{[16]}
I. Balitskii and A. Yung, {\it Phys. Lett.} {\bf B168}
(1986) 113;
A. Yung,
{\it Nucl. Phys.} {\bf B297} (1988) 47.
\item{[17]}
L. Yaffe, in {\it Proc. of the Santa Fe Workshop on Baryon Number
Violation at the SSC?}, eds. M. Mattis and E. Mottola
(World Scientific, Singapore, 1990);
P. Arnold and M. Mattis, {\it Phys. Rev.} {\bf D42} (1990) 1738.
\item{[18]}
S. Khlebnikov, V. Rubakov and P. Tinyakov,
{\it Nucl. Phys.} {\bf B350} (1991) 441.
\item{[19]}
A.H. Mueller,
{\it Nucl. Phys.} {\bf B348}
 (1991) 310 and {\bf B353} (1991) 44;
M. Voloshin,
{\it Nucl. Phys.} {\bf B359} (1991) 301.
\item{[20]}
 V. Zakharov, Minnesota preprint
 TPI-MINN-90/7-T (1990); {\it Nucl. Phys.} {\bf B371} (1992) 637.
\item{[21]}
M. Porrati, {\it Nucl. Phys.} {\bf B347} (1990) 371.
\item{[22]}
V.V. Khoze and     A. Ringwald,
{\it Nucl. Phys.} {\bf B355} (1991) 351.
\item{[23]}
D. Diakonov and V. Petrov, in {\it
Proc. of  the 26th Winter School
of the Leningrad Nuclear Physics Institute} (1991);
A.H. Mueller,
{\it Nucl. Phys.} {\bf B364} (1991) 109;
P. Arnold and M. Mattis,
{\it Mod. Phys. Lett.} {\bf A6} (1991) 2059.
\item{[24]}
D. Diakonov and M. Polyakov, St. Petersburg preprint
LNPI-1737 (1991).
\item{[25]} I. Balitskii and A. Sch\"afer, Frankfurt preprint UFTP
323 (1992).
\item{[26]} P. Silvestrov, Novosibirsk preprint BUDKERINP 92 (1992).
\item{[27]} M. Mattis, {\it Phys. Rep.} {\bf 214} (1992) 159;
P. Tinyakov, CERN preprint CERN-TH.6708 (1992).
\item{[28]}
V.V. Khoze   and    A. Ringwald,
{\it Phys. Lett.} {\bf B259} (1991) 106.
\item{[29]}
V. Zakharov, {\it Nucl. Phys.} {\bf B353} (1991) 683;
M. Maggiore and M. Shifman, {\it Phys. Rev.} {\bf D42} (1992) 3550;
G. Veneziano,
{\it Mod. Phys. Lett.} {\bf A7} (1992) 1661.
\item{[30]} H. Goldberg, {\it Phys. Rev.} {\bf D45} (1992) 2945.
\item{[31]} M. Voloshin, {\it Nucl. Phys.} {\bf B383} (1992) 233.
\item{[32]} E. Argyres, R. Kleiss and C. Papadopoulos,
{\it Nucl. Phys.} {\bf B391} (1993) 42.
\item{[33]} E. Argyres, R. Kleiss and C. Papadopoulos, {\it these
proceedings}.
\item{[34]} V. Zakharov, {\it Phys. Rev. Lett.} {\bf 67} (1991) 3650;
{\it Nucl. Phys.} {\bf B377} (1992) 501;
M. Maggiore and M. Shifman, {\it Nucl. Phys.} {\bf B380} (1992) 22.
\item{[35]} H. Goldberg, {\it Phys. Rev. Lett.} {\bf 69}
(1992) 3017.
\item{[36]} A. Ringwald, F. Schrempp and C. Wetterich,
{\it Nucl. Phys.} {\bf B365} (1991) 3.
\item{[37]} G. Farrar and R. Meng, {\it Phys. Rev. Lett.}
{\bf 65} (1990) 3377.
\item{[38]} D. Morris and R. Rosenfeld, {\it Phys. Rev.}
{\bf D44} (1991) 3530.
\item{[39]} D. Morris and A. Ringwald, CERN preprint
CERN-TH.6822/93 (1993).
\item{[40]} L. Bergstr\"om, R. Liotta and H. Rubinstein,
{\it Phys. Lett.} {\bf B276} (1992) 231;
L. Dell'Agnello et al., INFN
Firenze preprint DFF 178/12 (1992).
\item{[41]} D. Morris and A. Ringwald: ``Multiple Muons from
Neutrino-Initiated Multi-W(Z) Production", subm. to
{\it ICRC}, Calgary, Canada (1993).
\item{[42]} H. Meyer (Frejus Collaboration), in:
Proc. of the XXVIIth Recontre de Moriond, Les Arcs, Gif-sur-Yvette,
Ed. Frontieres, p. 169 (1992);
W. Allison et al. (Soudan 2 Collaboration), Argonne preprint
ANL-HEP-CP-92-39 (1992).
\item{[43]} F. Stecker et al., {\it Phys. Rev. Lett.} {\bf 66}
(1991) 2697; {\it ibid.} {\bf 69} (1992) 2738 (Erratum); for a review:
V. Stenger, DUMAND preprint DUMAND-9-92 (1992).
\item{[44]} J. Kempa and J. Wdowczyk, {\it J. Phys.} {\bf G9}
(1983) 1271; C. Forti et al., {\it Phys. Rev.} {\bf D42} (1990) 3668.
\item{[45]} R. Baltrusaitis et al. (Fly's Eye Collaboration),
{\it Phys. Rev.} {\bf D31} (1985) 2192.
\item{[46]} J. MacGibbon and R. Brandenberger,
{\it Nucl. Phys.} {\bf B331} (1990) 153.

\bye